\documentclass[10pt,conference]{IEEEtran}

\ifCLASSINFOpdf
\else
\fi
\usepackage{amsmath}
\usepackage{amssymb}
\usepackage{amsthm} 
\usepackage{booktabs}   
\usepackage{subcaption} 
\usepackage{tikz}
\usepackage{physics}

\usepackage{float} 

\usetikzlibrary{decorations.pathmorphing,shapes,decorations.markings}
\usepackage[all]{xy} 
\newdir{ (}{{}*!/-5pt/\dir^{(}}
\input{Qcircuit}
\usepackage{wrapfig}
\usepackage{tikz-cd}
\usepackage{todonotes}
\usetikzlibrary{positioning}
\newcommand\subfac{\mathord{\text{\textexclamdown}}}
\newcommand{\oone}{I}
\newcommand{\ozero}{N}
\newcommand{\RingoidTwo}[8]{\mathcal B(\begin{smallmatrix}#1 &#2& #3 &#4\\#5& #6 &#7& #8\end{smallmatrix})}

\newcommand{\TopRingoidTwo}[8]{\mathfrak{Top}\text{-}\mathcal{B}(\begin{smallmatrix}#1 &#2& #3 &#4\\#5& #6 &#7& #8\end{smallmatrix})}

\newcommand{\scatC}{\ensuremath{\mathbf{C}}} 

\newcommand{\scatD}{\ensuremath{\mathbf{D}}} 
\newcommand{\catV}{\ensuremath{\mathcal{V}}}
\newcommand{\Isometry}{\mathbf{Isometry}}
\newcommand{\Unitary}{\mathbf{Unitary}}
\newcommand{\Set}{\mathbf{Set}}
\newcommand{\Cstar}{\mathbf{Cstar}}
\newcommand{\CCstar}{\mathbf{CCstar}}
\newcommand{\CPTP}{\mathbf{CPTP}}
\newcommand{\CPU}{\mathbf{CPU}}
\newcommand{\Topo}{\mathfrak{Top}}

\newcommand{\CC}{\mathbb{C}}
\newcommand{\Embedding}{\mathbb{E}}
\newcommand{\id}{\mathrm{id}}
\newcommand{\Id}{\mathrm{Id}}
\newcommand{\Tra}{\mathrm{Tr}}
\newcommand{\Ad}{\mathrm{Ad}}
\newcommand{\MMet}{\mathbf{Met}}
\newcommand{\CMet}{\mathbf{CMet}}
\newcommand{\RR}{\mathbb R}
\newcommand{\QQ}{\mathbb Q}
\newcommand{\UU}{\mathbb U}

\newcommand{\defeq}{\stackrel{\text{def}}=}

\newcommand{\opp}{^{\mathrm{op}}}

\makeatletter
\let\save@mathaccent\mathaccent
\newcommand*\if@single[3]{%
  \setbox0\hbox{${\mathaccent"0362{#1}}^H$}%
  \setbox2\hbox{${\mathaccent"0362{\kern0pt#1}}^H$}%
  \ifdim\ht0=\ht2 #3\else #2\fi
  }
\newcommand*\rel@kern[1]{\kern#1\dimexpr\macc@kerna}
\newcommand*\widebar[1]{\@ifnextchar^{{\wide@bar{#1}{0}}}{\wide@bar{#1}{1}}}
\newcommand*\wide@bar[2]{\if@single{#1}{\wide@bar@{#1}{#2}{1}}{\wide@bar@{#1}{#2}{2}}}
\newcommand*\wide@bar@[3]{%
  \begingroup
  \def\mathaccent##1##2{%
    \let\mathaccent\save@mathaccent
    \if#32 \let\macc@nucleus\first@char \fi
    \setbox\z@\hbox{$\macc@style{\macc@nucleus}_{}$}%
    \setbox\tw@\hbox{$\macc@style{\macc@nucleus}{}_{}$}%
    \dimen@\wd\tw@
    \advance\dimen@-\wd\z@
    \divide\dimen@ 3
    \@tempdima\wd\tw@
    \advance\@tempdima-\scriptspace
    \divide\@tempdima 10
    \advance\dimen@-\@tempdima
    \ifdim\dimen@>\z@ \dimen@0pt\fi
    \rel@kern{0.6}\kern-\dimen@
    \if#31
      \overline{\rel@kern{-0.6}\kern\dimen@\macc@nucleus\rel@kern{0.4}\kern\dimen@}%
      \advance\dimen@0.4\dimexpr\macc@kerna
      \let\final@kern#2%
      \ifdim\dimen@<\z@ \let\final@kern1\fi
      \if\final@kern1 \kern-\dimen@\fi
    \else
      \overline{\rel@kern{-0.6}\kern\dimen@#1}%
    \fi
  }%
  \macc@depth\@ne
  \let\math@bgroup\@empty \let\math@egroup\macc@set@skewchar
  \mathsurround\z@ \frozen@everymath{\mathgroup\macc@group\relax}%
  \macc@set@skewchar\relax
  \let\mathaccentV\macc@nested@a
  \if#31
    \macc@nested@a\relax111{#1}%
  \else
    \def\gobble@till@marker##1\endmarker{}%
    \futurelet\first@char\gobble@till@marker#1\endmarker
    \ifcat\noexpand\first@char A\else
      \def\first@char{}%
    \fi
    \macc@nested@a\relax111{\first@char}%
  \fi
  \endgroup
}

\usepackage{circuitikz}
\usetikzlibrary{arrows,decorations.pathreplacing}
\tikzset{meter/.append style={draw, inner sep=10, rectangle, font=\vphantom{A}, minimum width=30, line width=.2,
 path picture={\draw[black] ([shift={(.1,.3)}]path picture bounding box.south west) to[bend left=50] ([shift={(-.1,.3)}]path picture bounding box.south east);\draw[black,-latex] ([shift={(0,.1)}]path picture bounding box.south) -- ([shift={(.3,-.1)}]path picture bounding box.north);}}}

\usepackage{thmtools,thm-restate}
\declaretheorem[style=plain,name=Theorem,numberwithin=section]{thm}
\newtheorem{theorem}[thm]{Theorem}
\newtheorem*{theorem*}{Theorem}
\newtheorem{proposition}[thm]{Proposition}
\newtheorem{corollary}[thm]{Corollary}
\newtheorem{lemma}[thm]{Lemma}

\theoremstyle{definition}
\newtheorem{definition}[thm]{Definition}
\newtheorem{example}[thm]{Example}

\theoremstyle{remark}
\newtheorem{remark}[thm]{Remark}
\newtheorem{notation}[thm]{Notation}

\newcommand{\mh}[1]{[{\color{blue} \textbf{MH: }#1}]}
\newcommand{\hide}[1]{}

\begin{document}
%
\title{Quantum channels as a categorical completion}

\author{
\IEEEauthorblockA{Mathieu Huot\\University of Oxford, UK}
\and
\IEEEauthorblockA{Sam Staton\\University of Oxford, UK}
}


\maketitle

\begin{abstract} 
We propose a categorical foundation for the connection between pure and mixed states in quantum information and quantum computation. The foundation is based on distributive monoidal categories. 

First, we prove that the category of all quantum channels is a canonical completion of the category of pure quantum operations (with ancilla preparations). More precisely, we prove that the category of completely positive trace-preserving maps between finite-dimensional C*-algebras is a canonical completion of the category of finite-dimensional vector spaces and isometries. 

Second, we extend our result to give a foundation to the topological relationships between quantum channels. We do this by generalizing our categorical foundation to the topologically-enriched setting. In particular, we show that the operator norm topology on quantum channels is the canonical topology induced by the norm topology on isometries.
\end{abstract}

%
\IEEEpeerreviewmaketitle

\section{Introduction}

 A popular explanation of quantum theory says that, in reality, everything is reversible (``pure quantum''), but conceptually we can hide and prepare things, and this is what leads to classical data, randomness and perceived irreversibility (``full quantum'').  In this paper we explain the passage from theories of pure quantum to theories of full quantum in terms of categorical completions. 

We test this passage in several ways:
\begin{itemize}
\item Starting from pure quantum with preparations (isometries), we recover quantum channels (completely positive maps between C*-algebras) as a completion with hiding --- this is our main result (Thm.~\ref{thm:main});
\item Starting from pure quantum (unitaries), we recover preparation of ancillas (isometries) as a completion with preparations (Thm.~\ref{thm:unitary});
\item Also starting from pure quantum (unitaries), we recover finite non-commutative geometry (finite-dimensional C*-algebras and $*$-homomorphisms) as a different completion (Thm.~\ref{thm:homo});
\item Starting from topologies on the isometries, we recover topologies on quantum channels as a completion (Thm.~\ref{thm:homoEnriched}).
\end{itemize}
All these require slightly different kinds of completion, and in this introduction we discuss the kinds of categories and completion at hand. First we consider the pure situation (\S\ref{sec:reversible}), then
preparation of states (\S\ref{sec:preparation}), and finally hiding of states (\S\ref{sec:hiding})
and topology (\S\ref{sec:topology}).
In what follows we use categorical terminology, but the casual reader may prefer the following informal picture of our main result.

\begin{center}
\begin{tikzpicture}[scale=1.1, decoration=snake]
   \draw [thick] (0,0) ellipse (3cm and 1.6cm);
   \draw [thick] (2,0) circle (1);
   \draw [->,>=stealth, decorate,  decoration={amplitude=0.7pt}] (-2,0) to[out=40] (2.47,0.58);
   \draw [thick,->,>=stealth] (-2,0)  -- (0.92,0) ;
   \draw [thick,->,>=stealth,dashed] (1,0) -- (2.42,0.5) ; 
   \node[] at (-0.5,-1) {};
	\node[align=left] at (2,-0.3) {\it admit \\ \it hiding};
	\node[draw,shape=circle,fill=black,minimum size=6pt,inner sep=0pt] at (-2,0) {};
	\node[below,align=left] at (-1.7,0) {\small Pure quantum \\[-2pt] \small + preparations};
	\node[draw,shape=circle,fill=black,minimum size=6pt,inner sep=0pt] at (1,0) {};
	\node[above,align=right] at (0.45,0) {\small Quantum \\[-2pt] \small channels};
	\node[draw,shape=circle,fill=black,minimum size=6pt,inner sep=0pt] at (2.5,0.5) {};
	\node[] at (0,1.3) {$\forall$};
	\node[] at (1.75,0.5) {$\exists!$};
\end{tikzpicture}
\end{center}

Informally, the outer ellipse contains all the possible theories, including pure quantum theory with preparations. The inner circle contains the theories that admit hiding. Our main result is that of all the theories that admit hiding, quantum channels are the `closest' to pure quantum with preparations. This notion of `closeness' will be made precise using category theory.

In \cite{huotuniversal} we presented a similar paradigm for the restricted version of quantum channels between matrix algebras. We proved that those quantum channels are the affine completion of the category of isometries, both seen as monoidal categories. We go further here by considering all finite dimensional C*-algebras which amounts to handling classical data. 

\subsection{Rudiments of pure / reversible computing}
\label{sec:reversible}
Before moving to categorical side, we recall some rudiments of reversible computing, which is one perspective on pure quantum theory. 
The basic idea is that a classical reversible operation on an $n$-level system is a bijection $n\to n$ on the natural number $n$ considered as a finite set. 
A \emph{quantum} reversible operation is an $n\times n$ complex matrix that is unitary. But the reader unfamiliar with quantum theory can focus on the classical setting for now, because every bijection can be thought of as a unitary matrix valued in $\{0,1\}$.
For example, there are two reversible classical operations on bits $2\to 2$, identity and negation, 
and a reversible 2-bit operation is a bijection $4\to 4$.  
The natural numbers form a rig (aka semiring) under addition and multiplication, and we find a simple calculus for building reversible operations by noticing that the bijections and unitaries can be composed but also they can be combined according to these rig operations. Here we write $(\oplus,N)$ and $(\otimes, I)$ instead of $+$ and $\times$ to emphasise their categorical nature.
\begin{itemize}
\item The multiplication of numbers corresponds to spatial juxtaposition of systems. For example, given two bijections on a bit, $f,g:2\to 2$, we have a bijection $(f\otimes g):2\otimes 2\to 2\otimes 2$ on $2\otimes 2\defeq 4$ on two bits. 
In terms of matrices, this is the Kronecker product.
\item The addition of numbers allows for conditional operations. Recall that most of the traditional logical operations are not reversible, however, it is possible to perform reversible controlled operations if the condition is kept. In terms of matrices, this is the block diagonal matrix. For example, the controlled-not gate 
is a bijection ${(\id\oplus \neg)}:{2\oplus 2}\to {2\oplus 2}$. Generally, given two unitaries ${f,g:n\to n}$, we can build a unitary ${(f\oplus g)}:{n\oplus n}\to {n\oplus n}$. Since $n\oplus n=2\otimes n$, we can think of $(f\oplus g):2\otimes n\to 2\otimes n$ as 
an operation that will either apply $f$ or $g$ to $n$ depending on the state of the first qubit, which is retained. 
\end{itemize}
The unit $I=1$ represents a system with no levels. There is only one classical bijection $1\to 1$, but in quantum computation, the unitaries $1\to 1$ correspond to angles in the interval $[0,2\pi)$, known as `global phase'. Here the addition plays a further role, since the unitary
${(0\oplus \frac \pi 4)}:2\to 2$ is called the $T$-gate. The controlled gates can be used to induce quantum entanglement in the product $2\otimes 2$. 
For example, consider the following circuit, which is a quantum Fourier transform on three qubits. 
It is a graphical notation for a unitary $2\otimes 2\otimes 2\to 2\otimes 2\otimes 2$. 
Vertical juxtaposition is $\otimes$; horizontal juxtaposition is composition of unitaries. The 
first gate $H:2\to 2$ is the Hadamard unitary gate, the next is a controlled $T$ gate
${(\id\oplus T)}:2\otimes 2\to 2\otimes 2$; the final gate is the swap gate $2\otimes 2\to 2\otimes 2$
which amounts to the symmetry of~$\otimes$. 
\begin{equation}\label{dgm:fft}\scalebox{0.8}{
\Qcircuit @C=.5em @R=.7em {
\ket {j_0} &&& \qw & \gate H& \gate T &\gate T &\gate T &\qw&\qw&\qw&\qw & \qw & \qswap\qw & \qw & \qw
\\
\ket {j_1} &&& \qw & \qw & \control{-1}{\qwx[-1]} \qw & \control{-1}{\qwx[-1]} \qw & \qw & \gate H & \gate T & \gate T & \qw & \qw &\qw & \qw & \qw
\\
\ket {j_2} &&& \qw & \qw & \qw & \qw & \control{-2}{\qwx[-2]} \qw &\qw & \control{-1}{\qwx[-1]} \qw &\control{-1}{\qwx[-1]} \qw & \gate H & \qw & \qswap \qwx[-2]\qw & \qw & \qw
}}
\end{equation}
We could thus equally well describe this circuit just by combining unitaries using the operations $\otimes,\oplus,\circ$. (For more details on this circuit see \cite{Nielsen:2011:QCQ:1972505}, Ch.~5.)

None of the above calculus \emph{requires} that the operations in question are reversible, rather, the point is that it applies even when the operations are reversible. 
Overall the structure of the calculus $(\otimes,\oplus,\circ)$ is that of a rig-category, in its strict form known as a bipermutative category (e.g.~\cite{laplaza,may1977ring,green-altenkirch}). 
This is a category with two monoidal structures $(\otimes ,I)$ and $(\oplus, N)$ which distribute over each other appropriately. 

\subsection{Preparation and initial objects}
\label{sec:preparation}
There are no bijections or unitaries $1\to 2$. However, it is very useful to be able to prepare a bit or qubit from an empty system. Thus we are led to consider injections between sets or isometries between vector spaces. Both of these structures again form a bipermutative category. 
There are two injections $1\to 2$, corresponding to the two possible states of a bit.
On the quantum side, these do not correspond to unitary matrices, but rather isometries.
There are uncountably many isometries $1\to 2$, obtained by composing the two injections $1\to 2$ with unitaries on $2$. These isometries describe the possible pure states of a qubit.

Injections (and isometries) again support a bipermutative category structure. In this setting, the zero system $N$, the unit for $\oplus$, can be thought of as an absurd uninhabited system, meaning that it is initial:
there is a unique map $\subfac:N\to A$ for all objects $A$. This induces the canonical basis injections ${I\xrightarrow =I\oplus N\xrightarrow{I\oplus \subfac} I\oplus I}$ in every bipermutative category with initial~$N$. 

Since preparation is arguably a conceptual abstraction in quantum theory, and not `real', it is reasonable to add it as freely as possible to the pure quantum theory of the unitaries. This is exactly what the isometries are, and this freeness is captured by the following theorem. Here we write $\Unitary$ and $\Isometry$ for the bipermutative categories of unitaries and isometries respectively. A `bipermutative functor' is a functor
that preserves all the bipermutative structure.
\begin{theorem*}[\ref{thm:unitary}]
\label{thm:isometry}
For every bipermutative category $\scatC$ with initial~$N$ 
and every bipermutative functor $F:\Unitary \to \scatC$,
there is a unique bipermutative functor 
$\hat F:\Isometry \to \scatC$ that makes the following diagram commute:
\begin{equation}\label{dgm:isom}
\begin{tikzcd}
	\Unitary \ar[r,hook,""] \ar[dr,"\forall F"']
	& \Isometry \ar[d,dotted,"\exists!\widehat{F}"] \\
	& \forall\scatC
\end{tikzcd}
\end{equation}
\end{theorem*}
This theorem is a standard kind of universal property. For a more familiar example, recall the property of 
the reals $\RR$ as a completion of the rationals $\QQ$: for any complete metric space
$S$ and any short map $f\colon \QQ\to S$ there is a unique short map $\hat f\colon \RR\to S$ such that 
\begin{equation}\label{eqn:realComp}
\begin{tikzcd}
	\QQ \ar[r,hook] \ar[dr,"\forall f"']
	& \RR \ar[d,dotted,"\exists!\widehat{f}"] \\
	& \forall S
\end{tikzcd}
\end{equation}
However, in \eqref{dgm:isom}, the objects ($\Unitary$ etc.) are themselves bipermutative categories rather than spaces, and the diagram is a diagram in the category of bipermutative categories. 
Nonetheless, as is usual, the universal property uniquely determines the bipermutative category $\Isometry$ up to isomorphism.

\subsection{Hiding, terminal objects, coproducts and classical data}
\label{sec:hiding}
In many irreversible situations, there is a unique map from any system to the empty system $I$. This is the operation of hiding (aka discarding) the state of a system. In categorical terms, $I$ is terminal. 
This is not the case in the categories of bijections, unitaries, injections or isometries, but it is the case in the category of sets and all functions, and also in the category of quantum channels. 
When $I$ is terminal, it is a general fact that the additive structure $(\oplus,N)$ of a bipermutative category is necessarily a categorical coproduct. This means that the object $2\defeq I\oplus I$ behaves like an object of classical bits, and indeed all the classical logic gates necessarily arise as morphisms $2^{\otimes n}\to 2$. 
This is in stark contrast to the reversible situation.

As with preparation, hiding is arguably an abstraction rather than `reality', and so it is reasonable to add it as freely as possible to pure quantum theories. This requires some care, as we now explain. 

The basic idea is to move from pure, semi-reversible situations to full quantum channels with classical data by freely turning $I$ into a terminal object.
There is a well-known functor \[\Embedding:\Isometry \to \CPTP\] 
and we will show that it has a universal property.
Here, $\CPTP$ is the category of quantum channels: finite dimensional C*-algebras as objects, and completely positive trace preserving maps as morphisms (\S\ref{subsec:MQT}). 
For now, we do not presume these definitions, and instead focus on the universal property. 
Some care is needed: if we formulated a universal property analogously to (\ref{dgm:isom}), we would lose all the quantum structure, because the additive quantum construction $\oplus$ would be collapsed into a categorical coproduct. 
For example, when $1$ becomes terminal, the global phases $1\to 1$ are all collapsed, which is desirable, but then the $T$ gate $(\id\oplus {\frac \pi 4})$ would also be collapsed with the identity gate, which is unacceptable. 
To avoid this, we do not ask for a completion that \emph{strictly} preserves the additive $\oplus$ structure, but only that there are comparison maps e.g.~$\Embedding(A\oplus B)\to \Embedding(A)\oplus \Embedding(B)$, in other words, that the embedding be \emph{colax} with respect to $\oplus$. 
For example, the comparison map $\Embedding(I\oplus I)\to \Embedding(I)\oplus \Embedding(I)$ takes a qubit to a classical bit, and corresponds to the fundamental operation of standard basis measurement. 

In circuit notation, the measurement is notated 
\scalebox{0.25}{\begin{circuitikz}[scale=1, transform shape]
\node[meter] (meter) at (0,0) {};
\draw(-1,0) -- (meter.west);
\draw(meter.07)  -- ++(.6,0);
\draw(meter.-07) -- ++(.6,0);	
\end{circuitikz}}
with the double-wire indicating a classical wire. For example, the 
following phase estimation circuit is an extension of~(\ref{dgm:fft}):
\[
\includegraphics[scale=0.7]{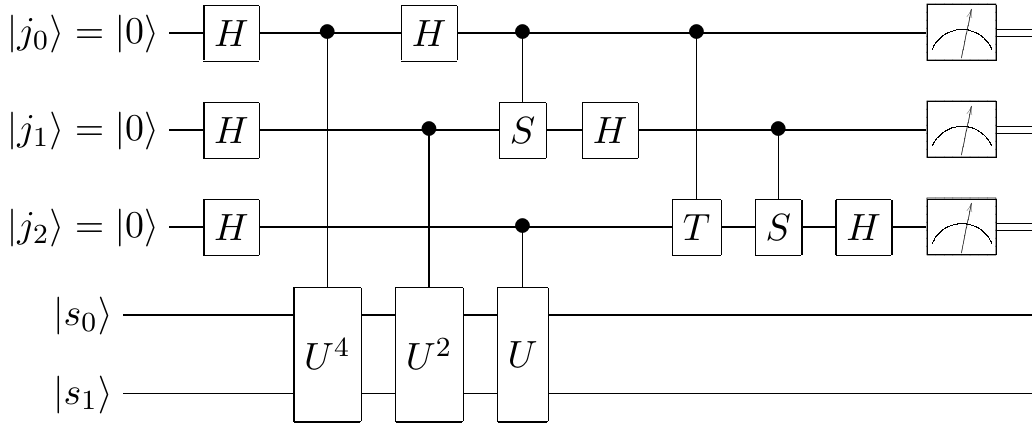}
\]
and describes a quantum channel \begin{multline*}
\Embedding(I\oplus I)^{\otimes 2}\xrightarrow{\text{preparation}  }
\Embedding(I\oplus I)^{\otimes 5}\xrightarrow{\text{unitary}}
\Embedding(I\oplus I)^{\otimes 5}\\\xrightarrow{\text{ measurement}}
(\Embedding(I)\oplus \Embedding(I))^{\otimes 3}\otimes \Embedding(I\oplus I)^{\otimes 2}\end{multline*}
taking two qubits to three classical bits and two qubits. (For more details on the circuit see e.g. \cite{Nielsen:2011:QCQ:1972505}, Ch.~5.)

This functor $\Embedding$ has the following universal property, which is the main result of this paper.
Here, we write `colax bipermutative functor' for a functor 
that strictly preserves $(\otimes,I)$ but is colax for $\oplus$, as $\Embedding$ is. 
\begin{theorem*}[\ref{thm:main}]
For every bipermutative category $\scatD$ with terminal~$I$ 
and every colax bipermutative functor ${F:\Isometry \to \scatD}$
there is a unique (strict) bipermutative functor 
$\hat F:\CPTP \to \scatD$ that makes the following diagram commute:
\begin{equation}\label{dgm:cptp}
\xymatrix{
\Isometry \ar[r]^-{\Embedding} \ar[dr]_{\forall F}
& \CPTP \ar@{.>}[d]^{\exists!\widehat{F}}
\\
& \forall\scatD
}
\end{equation}
 \end{theorem*}
Thus the category $\CPTP$ is the completion of $\Isometry$ with hiding.

As a step towards this theorem, we also consider a similar result for unitaries (Thm.~\ref{thm:homo}), for which the completion is a category of non-commutative spaces (C*-algebras and *-homomorphisms). 

\subsection{Topology quantum channels}
\label{sec:topology}
Classical computation is discrete, but many results in quantum computation and information are concerned with continuous aspects of quantum channels. 
For example, 
\begin{itemize}
\item \emph{circuit approximation:} for any quantum channel $F$ over qubits there is a sequence of quantum channels $F_1,F_2\dots$ that converges to $F$ such that each approximant $F_i$ is only built from standard basis measurement, preparations, and the $H$, $T$ and controlled-not gates (e.g.~\cite{Nielsen:2011:QCQ:1972505}, \S4.5). 
\item \emph{noisy quantum channels}: in many situations it is desirable for a quantum channel $F$ to leave a part of the system untouched: thus we are interested in the distance between the channel~$F$ and the identity channel (e.g.~\cite{Nielsen:2011:QCQ:1972505}, \S9.3). This may be non-zero because of imperfections in the circuit manufacture.
\item \emph{quantum topology}: much inspiration for understanding quantum theory is by studying the topology of the unitary groups (e.g.~\cite{kauffman-baadhio}). On a related note, a simple topological way of understanding the mixed states of a single qubit is as homeomorphic to the 3-ball (Bloch sphere). 
\end{itemize}
All these considerations require a notion of topology on the quantum channels themselves.

Our final theorem is that the operator norm topology on quantum channels is canonically induced by the spectral topology on the isometries. 
To be precise, recall that for every pair of numbers $m$, $n$, the commuting diagram~\eqref{dgm:cptp} of functors given in the universal property of $\CPTP$ induces in particular a commuting diagram of functions between hom-sets 
\begin{equation}\label{eqn:hom-functions}
\xymatrix{
\Isometry(m,n) \ar[r]^-{\Embedding_{m,n}} \ar[dr]_{ F_{m,n}}
& \CPTP(\Embedding(m),\Embedding(n)) \ar[d]^{\widehat{F}_{m,n}}
\\
& \scatD(m,n)
}
 \end{equation}
We can regard these hom-sets not just as sets but as topological spaces, and to show that the topology on quantum channels is canonical, we characterize  when these functions are continuous. 
\begin{theorem*}[\ref{thm:cptpEnriched}, paraphrased]
In diagram~\eqref{eqn:hom-functions}: ${\Embedding_{m,n}}$ 
is continuous, and if the hom-sets of $\scatD$ are equipped with a topology that respects the bipermutative category structure and such that 
$F_{m,n}$ 
is continuous, then 
$\hat F_{m,n}$ 
is also continuous.
\end{theorem*}
The usual way to understand categories whose hom-sets have structure is by `enriched' category theory, 
and this is how we proceed in Section~\ref{sec:enriched}. 

\medskip
\paragraph*{Summary}
We have highlighted how the framework of bipermutative categories (\S\ref{sec:reversible}) can be used to give universal properties to theories with preparation (\S\ref{sec:preparation}), preparation and hiding (\S\ref{sec:hiding}), and to the topology of quantum channels (\S\ref{sec:topology}). 

We now proceed to elaborate these matters, in Sections~\ref{sec:prel}, \ref{subsec:C}, \ref{sec:cptp} and \ref{sec:enriched} respectively.

 
\section{Bipermutative categories as models of QT}
\label{sec:prel}

In our introductory remarks~(\S\ref{sec:reversible}), we discussed the role that bipermutative categories play in a general form of circuit-like structure that is relevant even in the reversible setting. 
We now recall the key definitions (\S\ref{subsec:GS}) and explain how variations on quantum theory can be understood as bipermutative categories (\S\ref{subsec:MQT}). 

\subsection{Rig categories and bipermutative categories}
\label{subsec:GS}

We begin by recalling the notion of rig category, before discussing the stricter variant of bipermutative categories.  

\begin{definition}
  A \emph{symmetric monoidal category} is a tuple $(\scatC,\otimes,I,\gamma,\alpha,\lambda,\rho)$, where $\scatC$ is a category, $\otimes:\scatC\times\scatC\to\scatC$ is a functor, $\alpha_{A,B,C}:(A\otimes B)\otimes C\to A\otimes (B\otimes C)$, $\lambda_A:I\otimes A\to A$, $\rho_A:A\to A\otimes I$ and $\gamma_{A,B}:A\otimes B\to B\otimes A$ are natural isomorphisms. They are also required to satisfy some coherence conditions \cite{maclane:71}.

We will often omit $\alpha,\lambda,\rho$ when the context is clear or when they are identities.

A \emph{rig category} is a category with two symmetric monoidal structures, 
$(\oplus,N,\gamma)$ and $(\otimes,I,\gamma')$, such that one distributes over the other, that is, there are natural isomorphisms 
$\delta:(A\otimes B)\oplus (A\otimes C)\to A\otimes(B\oplus C)$ 
and 
$\delta^\sharp:(A\otimes C)\oplus (B\otimes C)\to (A\oplus B)\otimes C$,
$\lambda^*:N\to N\otimes A$ and 
$\rho^*:N\to A\otimes N$
satisfying some coherence conditions \cite{laplaza}. 
\end{definition}
\begin{example}
\begin{itemize}
	\item Any rig $R$ (aka~semiring: ring except no inverses for addition) can be seen as a rig category whose objects are the elements of $R$ and no non-identity morphisms, with the two monoidal structures given by the addition and multiplication of the rig. 
	\item The category of sets and functions where $\oplus$ is the disjoint union  and $\otimes$ the Cartesian product. 
        \item Also some variations: the category of finite sets and functions between them; 
          the category of finite sets and bijections between them; the category of finite sets and injections between them. 
        \item `Skeletal' variations on the above. e.g.  The category of natural numbers considered as sets $n=\{1\dots n\}$ and where morphisms are functions. In this example $m\oplus n=m+n$, $m\otimes n=mn$.
\end{itemize}
\end{example}
The 24 coherence conditions \cite{laplaza} are cumbersome, but they are simplified in many cases. 
We can often choose a strict monoidal structure and we can choose one of the distributivity isomorphisms to be an identity morphism, but the other distributivity isomorphism and the symmetries need not be identities. 

The skeleton category of the category of finite sets and bijections is such an example. Its symmetries are not identities. One of the distributivity isomorphisms can be identity, but the other cannot be. 

Every rig-category is equivalent to a bipermutative category (\cite{may1977ring}, Prop.~3.5).
\begin{definition}
A symmetric monoidal category $(\scatC,\otimes,I,\gamma,\alpha,\lambda,\rho)$ is \emph{strict} when $\alpha,\lambda,\rho$ are identity natural transformations.

A \emph{bipermutative category} is a rig category where both symmetric monoidal structures are strict, and $\delta$, $\lambda^*$ and $\rho^*$ are identity natural transformations
(but $\delta^\sharp$ need not be identity).
\end{definition}
Many of the coherence conditions for rig categories become trivial for bipermutative categories, because of the identity morphisms. Indeed $\delta^\sharp$ is uniquely determined by the symmetry $\gamma'$ in a bipermutative category.

\medskip

\subsubsection{Rig categories with coproducts}
In many rig categories, $\oplus$ is a coproduct. Indeed, 
in a symmetric monoidal category with coproducts $+$, one can always define canonical morphisms
\[
0\to A\otimes 0
\quad
(A \otimes B) + (A\otimes C)\to A \otimes (B+ C)
\]
and symmetrically. If these are all isomorphisms, then the coherence conditions for a rig category are automatically satisfied. 
The category of bijections is not of this form, and nor is the category of unitaries discussed below,
but the category of quantum channels does have coproducts.  

In fact this coproduct scenario is unavoidable when the multiplicative unit $I$ is terminal. 
This is a common assumption on $I$; in linear logic it is sometimes called affine logic and amounts to allowing discarding but not duplication of resources~\cite{jacobs-affine}. 
\begin{proposition}
Let $(\scatC,\oplus,N,\otimes,I)$ be a bipermutative category where $I$ is terminal. 
Then $(\oplus,N)$ is a categorical coproduct when considered with the following left injection
\[
A = A \oplus  (B \otimes N) \xrightarrow{A\oplus (B \otimes !)}  A\oplus (B\otimes I)\ =\ A \oplus B
\]
and similar right injection.
\end{proposition}
\begin{proof}[Proof sketch]
We could not find this result in the literature.
The unique copairing of ${A \xrightarrow f C \xleftarrow g B}$ is 
\[
{A\oplus B \xrightarrow{f\oplus g} C\oplus C \xrightarrow = (I\oplus I)\otimes C \xrightarrow {!\otimes C} I\otimes C\xrightarrow = C\text.}\]
The initial maps are $N \xrightarrow= {A \otimes N} \to A\otimes I\xrightarrow = A$.
\end{proof}

\subsubsection{Bipermutative functors and categories of bipermutative categories}
In this work we are interested in the relationships between variations on quantum theories, and these relationships will be described as bipermutative functors of certain kinds. 
\begin{definition}
A \emph{strict} bipermutative functor is a functor ${F:\scatC\to\scatD}$ between bipermutative categories $(\scatC,\oplus,N,\gamma,\otimes,I,\gamma')$ and $(\scatD,\oplus,N,\gamma,\otimes,I,\gamma')$ that strictly preserves all the structure: ${F(N)=N}$, $F{(I)=I}$, ${F(A\otimes B)}=F(A)\otimes F(B)$, $F(A\oplus B)=F(A)\oplus F(B)$, $F(\gamma_{A,B})=\gamma_{FA,FB}$, $F(\gamma_{A,B}')=\gamma_{FA,FB}'$.
\end{definition}
\begin{definition}
A $\oplus$-\emph{colax} bipermutative functor is a functor $F:\scatC\to\scatD$ between bipermutative categories such that $F(I)=I$, $F(A\otimes B)=F(A)\otimes F(B)$, $F(\gamma')=\gamma'$, together with natural transformations $\varphi:F(A\oplus B)\to F(A)\oplus F(B)$ and a morphism $\phi:F(N)\to N$ such that the following diagrams commute:
\[
\xymatrix{
F(A\oplus B)\otimes FC \ar[r]^{\varphi\otimes FC}
\ar@{=}[d]
& 
\ar@{=}[d]
(FA\oplus FB)\otimes FC
\\
F((A\otimes C)\oplus(B\otimes C))\ar[r]_\varphi 
& 
(FA\otimes FC)\oplus (FB\otimes FC)
}\]
\[
\xymatrix{
FN\otimes FA \ar[r]^{\psi \otimes FA}
\ar@{=}[d]
& 
\ar@{=}[d]
N\otimes FA
\\
FN \ar[r]_\psi
& 
N
}\]
together with the coherence diagrams making $(F,\varphi,\psi)$ a symmetric colax monoidal functor (\cite[XI.2]{maclane:71}, \cite{lewis1972coherence}) for $(\oplus,N)$.
\end{definition}
(When there are natural transformations ${F(A)\oplus F(B)}\to {F(A\oplus B)}$ instead, with similar coherence conditions, then $F$ is called a $\oplus$-\emph{lax} bipermutative functor, but we have less use for this notion.) 

\hide{To show that a model has a canonical status, we show it's in the way the closest among possible models. Hence we group those possible models in a category and hence we will consider categories whose objects are bipermutative categories and whose morphisms are either lax or strict bipermutative functors. 
We will sometimes restrict to bipermutative categories where $N$ is an initial object, or $I$ is a terminal object, or $\oplus$ is a coproduct, the latter case $\delta$, $\delta^\sharp$ and $\gamma$ should be the canonical morphisms. It reflects the idea that to add structure to a model, you consider a new model with this additional structure. There are candidate models with the right additional structure which form a category which is a subcategory of the category in which the first model lives.
  }

Strict/colax bipermutative functors compose, and so we can build categories of bipermutative categories and strict/colax bipermutative functors. 
Because there are various categories that we consider, we introduce the following notation. 
\begin{notation}
\label{ringoid_not}
The bipermutative functors that will be considered hereafter are required to strictly preserve the unit $N$ of $\oplus$, that is, $\psi=id_N$. We denote by
\begin{itemize}
	\item $\RingoidTwo\otimes I \oplus N ssss$ the category of bipermutative categories and strict bipermutative functors
	\item $\RingoidTwo\otimes I \oplus N sscs$ the category of bipermutative categories and $\oplus$-colax bipermutative functors
	\item $\RingoidTwo\otimes I \oplus 0 ssss$ the category of bipermutative categories for which the unit of $\oplus$ is an initial object, and strict bipermutative functors 
	\item $\RingoidTwo\otimes I \oplus 0 sscs$ the category of bipermutative categories for which the unit of $\oplus$ is an initial object, and $\oplus$-colax bipermutative functors 
	\item $\RingoidTwo\otimes 1 + 0 ssss$ the category of bipermutative categories for which the unit of $\otimes$ is a terminal object and $\oplus$ is a coproduct, and strict bipermutative functors 
\end{itemize}

\end{notation}
(By using bipermutative categories instead of rig categories, we are able to work with 1-categories such as $\RingoidTwo\otimes I\oplus N ssss$ instead of 2-categories and bicategories, which would be a distraction in this paper.)

\subsection{Models of Quantum Theory}
\label{subsec:MQT}

We can now express variations of quantum theory as bipermutative categories.

\begin{definition}
An $n\times n$ complex matrix $U$ is \emph{unitary} if
$U^*U=UU^*=I$, where $U^*$ is the conjugate transpose. The bipermutative category of unitaries is formed as follows. The objects are natural numbers (including zero). 
There is a morphism $n\to n$ for each $n\times n$ unitary and the set of $n\to n$ unitaries is usually denoted by $\UU(n)$.
\begin{itemize}
\item Composition of morphisms is matrix multiplication.
\item On objects, $\oplus$ is addition of numbers. On morphisms it leads to control gates. 
Given unitaries $U:n\to n$ and $V:m\to m$, we let $U\oplus V:n\oplus m\to n\oplus m$ be the block diagonal matrix $\begin{bmatrix}U&0\\0&V\end{bmatrix}$. 
The unit $N$ is the number zero.
\item On objects, $\otimes$ is multiplication of numbers. On morphisms, given unitaries
  $U:n\to n$ and $V:m\to m$, we let $U\otimes V:n\otimes m\to n\otimes m$ be the Kronecker product of 
  matrices, with $(U\otimes V)_{in+k,jn+l}=U_{i,j}V_{k,l}$. The unit $\oone$ is the number $1$.
\end{itemize}
\end{definition}

$\CC^2$ is to be thought as the state space of qubits. With this in mind, $\CC^{2^n}$ is the state space for $n$ qubits. Then the symmetry $\gamma_{2,2}':2\otimes 2\to 2\otimes2$ represents the swap gate and $\gamma_{1,1}:1\oplus1\to1\oplus1$ the not gate. Intuitively, $\id_n\otimes V:nm\to nm$ represents the transformation where $V$ acts on the subsystem $m$ of $nm$ without perturbing the subsystem $n$. Whereas $\otimes$ keeps separate (untangled) systems separate,
$\oplus$ is a source of entanglement as for instance $\id_2\oplus\gamma_{1,1}:4\to 4$ represents the controlled-not gate, which performs a non-trivial unitary transformation on the second qubit depending on the state of the first one. 

\begin{definition}
An \emph{isometry} is a linear map $\CC^m\to \CC^n$ which preserves the inner product metric. In other words, an isometry is an $n\times m$ complex matrix $V$ such that $V^*V=I$. Note that necessarily $m\leq n$ and $m=n$ precisely when an isometry is unitary. We form the bipermutative category of isometries in the same way as the one for unitaries.
The unit $N$ is the number zero; it is an initial object, as witnessed by the empty matrices $\subfac_n:0\to n$.
\end{definition}

\medskip
\subsubsection*{C*-algebras}
The idea of von Neumann's full quantum theory is that mixed states are understood as operators, for example, density matrices, or more generally elements of C*-algebras. 
Recall that a \emph{complex algebra} is a vector space $V$ over $\CC$ with an additional binary operation $\cdot:V\times V\to V$ that is linear in each argument.
The $n\times n$ complex matrices $\mathcal{M}_n(\CC)$ are an example of such an algebra where $\cdot$ is given by matrix multiplication. We also consider direct sums of matrix algebras $\bigoplus_{j\in J} \mathcal{M}_{n_j}(\CC)$ where $J$ is a finite set for the corresponding vector space which inherits the algebra structure componentwise. 

A complex $*$-algebra $A$ is a complex algebra with an additional operation $^*:A\to A$ such that ${\bf x}^{**}={\bf x}$, $(a{\bf x})^*=\bar{a}{\bf x}^*$, $({\bf x}+{\bf y})^*={\bf x}^*+{\bf y}^*$, $({\bf xy})^*={\bf y}^*{\bf x}^*$ and $1^*=1$ for all complex numbers $a$ and all ${\bf x},{\bf y}\in A$, where $\bar{a}$ is the complex conjugate of $a$. 

$\mathcal{M}_n(\CC)$ have a $*$-algebra structure given by the conjugate transpose and $\bigoplus_{j\in J} \mathcal{M}_{n_j}(\CC)$ again inherits $*$-algebra structure componentwise. Such algebras can be equipped with the spectral norm and they are complete with respect to this norm. These are finite dimensional C*-algebras and every finite dimensional C*-algebra is of this form, up to isomorphism. We will therefore from now on use `C*-algebra' to mean a finite dimensional C*-algebra of the form $\bigoplus_{j\in J} \mathcal{M}_{n_j}(\CC)$.

A positive element $A$ in a C*-algebra is such that there exists an element $B$ such that $A=B^*B$, or equivalently if it is self-adjoint and its spectrum $\sigma(A)$ consists of non-negative real numbers.

A linear map $f:A\to B$ between two C*-algebras is \emph{positive} if it maps positive elements to positive elements. 
A linear map $f:A\to B$ between two C*-algebras is \emph{completely positive} (CP) if for every $k$ the map $\id_{\mathcal{M}_k(\CC)}\otimes f:\mathcal{M}_k(\CC)\otimes A\to \mathcal{M}_k(\CC)\otimes B$ is positive. 

If we denote by $\Tra$ the trace operator, then a linear map $f:\mathcal{M}_n(\CC)\to \mathcal{M}_m(\CC)$ is said to be trace-preserving if for all $M\in \mathcal{M}_n(\CC)$, $\Tra(f(M))=\Tra(M)$. We can extend the trace operator to C*-algebras by $\Tra(A_1,\ldots,A_k)=\Tra(A_1)+\ldots \Tra(A_k)$ which is to say we embed $\bigoplus_{i\in I} \mathcal{M}_{n_i}(\CC)$ into $\mathcal{M}_{\sum_in_i}(\CC)$ as block diagonal matrices ($(A_1,\ldots,A_k)\mapsto A_1\oplus\ldots \oplus A_k$) and then apply the usual trace operator.

In von Neumann's model, a state is represented by a positive element of trace $1$.
For instance, a state of a qubit is a density matrix in $\mathcal M_2(\CC)$, and a state of a classical bit is a positive element of trace~$1$ in $\CC\oplus \CC$ (there are only two). 

A valid transformation of states is a linear map that must satisfy the following properties. First, it should send states to states, so it should send positive elements to positive elements and preserve trace. In addition, as it may act on a subsystem only, the total system should remain a state, hence a positive element. It means tensoring the map with the identity should also be a positive map and so the function is asked to be completely positive.
(The transpose map is a typical example of a positive but not completely positive map.)
Thus, completely positive trace preserving maps (CPTP) are sometimes called \emph{quantum channels}.

We note that for \emph{commutative} C*-algebras $\CC^m,\CC^n$, a CPTP map is the same thing as a stochastic $n\times m$ matrix, i.e.~a positive-real valued matrix where each column sums to $1$. 
(See e.g.~\cite{furber2013kleisli}.) This is a standard model of finite probabilistic classical computation and information, and so the non-commutative C*-algebras are a natural quantum generalization of this.

\begin{definition}
The bipermutative category of completely positive trace-preserving (CPTP) maps is defined as follows. Its objects are finite lists of positive natural numbers and a morphism $f:[n_1,\ldots,n_k]\to [m_1,\ldots,m_p]$ is a CPTP map $f:\bigoplus_i \mathcal{M}_{n_i}(\CC)\to \bigoplus_j \mathcal{M}_{m_j}(\CC)$. For instance $\mathcal{M}_2(\CC)$ is the state space of qubits and $\CC\oplus\CC$ is the space for bits.

$\oplus$ is given on objects by concatenation of lists and on morphisms $f:A\to B,g:A'\to B'$ by $(f\oplus g)(a,b):=\big(f(a),g(b)\big)$. $\oplus$ is a coproduct in this category. The empty list is the initial object.
$\otimes$ is given on objects by $[n_1,\ldots,n_k]\otimes [m_1,\ldots,m_p]:=[n_1m_1,\ldots,n_1m_p,\ldots,n_km_1,\ldots,n_km_p]$ and on morphisms $f:A\to [n],g:B\to [m]$ by $(f\otimes g):A\otimes B\to [nm]:(a\otimes b)\mapsto f(a)\otimes f(b)$ where $\otimes$ is the Kronecker product.
The object $[1]$, to be thought as $\CC$, is terminal and the terminal map $!_A:A\to [1]$ is given by the trace operator. The map $!_n\otimes \id_m:[nm]\to [m]$ is usually called the partial trace.
\end{definition}

\begin{definition}
We define an $\oplus$-colax-bipermutative functor $\Embedding:\Isometry\to\CPTP$ which sends $n \to [n]$ and $V:m\to n$ to $\Ad_V:M\mapsto VMV^*:[m]\to[n]$. The colax $\oplus$-morphism $\varphi_{n,m}:\mathcal{M}_{n+m}(\CC)\to\mathcal{M}_{n}(\CC)\oplus\mathcal{M}_{m}(\CC)$ is then defined by $\begin{bmatrix}A&B\\C&D\end{bmatrix}\mapsto(A,D)$.
\end{definition}

Intuitively, $\varphi_{2^n,2^n}:\mathcal{M}_{2^{n+1}}(\CC)\to(\CC\oplus\CC)\otimes \mathcal{M}_{2^n}(\CC)$ measures the first qubit of a system of $n+1$ qubits, so the resulting system consist of a bit and $n$ qubits. Those $n$ qubits are in the state $\Tra(A)^{-1}A$ if $0$ is measured, which happens with probability $\Tra(A)$, and $\Tra(D)^{-1}D$ otherwise.


\section{Isometries as a completion of unitaries}
\label{subsec:C}

We recall some basic category theory before characterizing the category of Isometries as a completion (Thm.~\ref{thm:unitary}).
In the introduction (\S\ref{sec:preparation}) we motivated universal completions by considering the inclusion $\QQ\hookrightarrow\RR$ of the rationals in the reals, which satisfies a universal property \eqref{eqn:realComp}. In general:
\newcommand{\sF}{\mathbf{F}}
\begin{definition}
If $\sF\colon{\scatC\to \scatD}$ is a functor and $e\colon {X\to \sF(Y)}$ is such that for every $f\colon {X\to \sF (Z)}$ there is a unique $\widehat{f}\colon {Y\to Z}$ such that the following diagram commutes:
\[
\xymatrix{
X \ar[r]^{e} \ar[dr]_{\forall f}
& \sF(Y) \ar@{.>}[d]^{\sF(\widehat{f})}
&Y\ar@{.>}[d]^{\exists!\widehat f}
\\
& \sF(Z)
& Z
}
\]then we say that $e:X\to \sF(Y)$ is a \emph{completion w.r.t.~$\sF$}.
(Other terminology: $e$ is universal for $F$; $e$ is free w.r.t.~$\sF$.)
\end{definition}

In particular the inclusion $\QQ\hookrightarrow \RR$ is a completion w.r.t. the forgetful functor $\CMet\to \MMet$, where $\MMet$ is the category of metric spaces, and $\CMet$ the category of complete metric spaces.

The main results of the paper are completions in this sense and we now give our first. 
The category of isometries is the initial object completion of the category of unitaries, both seen as bipermutative categories. 
The completion then may be read as saying that the category of isometries is the simplest model for pure quantum theory with ancillas (and no discarding).

The following proposition is the key point in the proof of the completion:

\begin{proposition}
\label{propn:caracIsometry}
If $V:m\to n$ is an isometry then
there is a unitary $U:n\to n$ such that
$V=U(I_m\oplus {\subfac_p})$, where $n=m+p$.

Moreover, $U$ is essentially unique: if $U_1,U_2:n\to n$ are such that $U_1(I_m\oplus {\subfac_p})=U_2(I_m\oplus {\subfac_p})$,
then there is a unique unitary $W:p\to p$ such that $U_2=U_1(I_m\oplus W)$. 
\end{proposition}

\begin{proof}
	Note that $(I_m\oplus {\subfac_p})$ is the isometry 
$\big(\begin{smallmatrix}I_m\\0\end{smallmatrix}\big):m\to m+p$,
and the requirement $V=U(I_m\oplus {\subfac_p})$ means that 
$U=(V|V')$ for some $n\times p$ matrix $V'$.

For the existence part, note that the columns of $V$ form a set of orthonormal vectors in $\CC^n$ which we can thus extend to an orthonormal basis for $\CC^n$, which forms the columns of a unitary $U$. One way to do this is by picking $p$ vectors in $\CC^n$ that are linearly independent of $V$, and then using the Gram-Schmidt process to turn this into a basis.

Now suppose we have two such $U_1$ and $U_2$. 
Note that $U_1$ and $U_2$ can be written as block matrices
$U_1=\big(\begin{smallmatrix}A&C_1\\B&D_1\end{smallmatrix}\big)$
and $U_2=\big(\begin{smallmatrix}A&C_2\\B&D_2\end{smallmatrix}\big)$,
where $A$ is $m\times m$, $B$ is $p\times m$, $C_1$ and $C_2$ are $m\times p$, and $D_1$ and $D_2$ are $p\times p$.
Note that $V=\big(\begin{smallmatrix}A\\B\end{smallmatrix}\big)$. 
Since $U_1$ and $U_2$ are unitaries, we have for $i=1,2$:
 \[I=U_i^*U_i=\begin{pmatrix}A^*A+B^*B&A^*C_i+B^*D_i\\C_i^*A+D_i^*B&C_i^*C_i+D_i^*D_i\end{pmatrix}
\] 
In particular, 
$A^*A+B^*B=I$, $C_1^*A+D_1^*B=0$, and $A^*C_2+B^*D_2=0$. 
Now we use these three facts to calculate 
\begin{align*}
	U_1^*U_2
	&=\begin{pmatrix}A^*A+B^*B&A^*C_2+B^*D_2\\C_1^*A+D_1^*B&C_1^*C_2+D_1^*D_2\end{pmatrix} \\
	&=\begin{pmatrix}I&0\\0&C_1^*C_2+D_1^*D_2\end{pmatrix}
\end{align*}
Let $W=C_1^*C_2+D_1^*D_2$. It is a unitary because its columns are orthonormal.
We have $(I\oplus W)=U_1^*U_2$, so $U_1(I\oplus W)=U_2$, as required.
\end{proof}

We can now state the theorem of the section:

\begin{theorem}
\label{thm:unitary}
The functor $\Unitary\to\Isometry$ is a completion w.r.t. the embedding 
$\RingoidTwo\otimes \oone \oplus 0 ssss
\to
\RingoidTwo\otimes \oone \oplus\ozero ssss$. That is:
\[
\begin{tikzcd}
	\Unitary \ar[r,hook,""] \ar[dr,"\forall F"']
	& \Isometry \ar[d,dotted,"\exists!\widehat{F}"] \\
	& \forall\scatC 
\end{tikzcd}
\]
where $F\in\RingoidTwo\otimes \oone \oplus\ozero ssss$ and $\widehat{F}\in\RingoidTwo\otimes \oone \oplus 0 ssss$. See notation \ref{ringoid_not}.
\end{theorem}

\section{Full QM without ancillas}
\label{sec:homo}

Unitaries are a model of pure quantum theory without ancillas. 
We are building up to a model of quantum theory \emph{with} ancillas in Section~\ref{sec:cptp}, 
but as a first step we consider what happens when we complete unitaries with hiding. As we show (Thm.~\ref{thm:homo}),
this results in the dual $\Cstar\opp$ of the bipermutative category of finite dimensional C*-algebras and *-homomorphisms.

If the reader is unfamiliar with $\Cstar\opp$, they may think of it as a `quantum' or `non-commutative' extension of the category of finite sets and functions. Indeed, the full subcategory $\CCstar\opp\subseteq \Cstar\opp$ of \emph{commutative} C*-algebras is equivalent to the category of finite sets and functions. This is because the finite-dimensional commutative C*-algebras are all of the form $\CC^n$, and every function 
${m\to n}$ induces a *-homomorphism ${\CC^n\to \CC^m}$ by reindexing, 
and every *-homomorphism arises in this way. 
(This is a starting point for `non-commutative geometry'.)

\hide{So it is expected that the completion of unitaries adding discarding will be a model of quantum theory without ancillas, a subcategory of $\CPTP$. 
We hence take a different approach and use duality. 
A known result on duality states that the dual of the category $\CPTP$ is the category of unital CP maps, \emph{i.e.} those CP maps that preserve the unit of the $\mathbb{C}^*$-algebra (the identity matrix in the case of a matrix algebra). 
$*$-homomorphisms are morphisms that respect the structure of $\mathbb{C}^*$-algebras.
The category $\CPTP$ has a non-full subcategory of unital $*$-homomorphisms. We show that the category of unital $*$-homomorphisms is a completion of the opposite of the category of unitaries and then obtain the main result of the section by duality.}

The key step in our proof of Theorem~\ref{thm:homo}, which gives a canonical status to $\Cstar\opp$, is a combinatorial characterization of the *-homomorphisms due to Bratteli~\cite{bratteli1972inductive}.

\subsection{Duality}

\begin{definition}
A linear map $\pi:A\to B$ between C*-algebras is called a $*$-homomorphism if:
\begin{itemize}
\item $\forall x,y\in A, \pi(x\cdot y)=\pi(x)\cdot\pi(y)$
\item $\forall x\in A,\pi(x^*)=\pi(x)^*$
\end{itemize}
It is unital if it preserves the unit of the C*-algebra.
\end{definition}

\begin{definition}
We have a bipermutative category $\Cstar$ of unital $*$-homomorphisms. Its objects are finite lists of positive natural numbers and a morphism $f:[n_1,\dots,n_k]\to [m_1,\dots,m_p]$ is a unital $*$-homomorphism $f:\bigoplus_{i}\mathcal{M}_{n_i}(\CC)\to \bigoplus_{j}\mathcal{M}_{m_j}(\CC)$.
$\oplus,\otimes$ are given similarly to those in $\CPTP$ (See \ref{subsec:MQT}).

Similarly, we define the bipermutative category $\CPU$ of completely positive unital maps. Its objects are those of $\Cstar$ and its morphisms are completely positive unital maps. $\oplus,\otimes$ are again given similarly to those in $\CPTP$.
\end{definition}

$\Cstar$ is a non-full subcategory of $\CPU$ and we have the following known result from Choi which is key to the duality:

\begin{proposition}(Choi \cite{choi1975completely})\label{prop:choi}
$f:\mathcal{M}_n(\mathbb{C})\to\mathcal{M}_m(\mathbb{C})$ is completely positive iff it admits an expression $f(A)=\sum_i V_i^*AV_i$ where $V_i$ are $n\times m$ matrices.
\end{proposition}

Recall that every linear map $A\to B$ between finite dimensional spaces has a dual $B\to A$. From Proposition~\ref{prop:choi}, a linear map $A\to B$ between C*-algebras is completely positive if and only if its dual $B\to A$ is completely positive. A completely positive linear map is furthermore trace preserving (CPTP) if its dual is furthermore unit preserving (CPU). 

\begin{corollary}
$(\CPU)^{op}\cong\CPTP$.
\end{corollary}

Furthermore, the functor $\Isometry\to\CPTP$ restricts to a functor $\Unitary\to\Cstar\opp$.

The following is an explicit characterisation of unital $*$-homomorphisms in finite dimension:

\begin{proposition}(Bratteli \cite{bratteli1972inductive})
$f:\bigoplus_{1\leq i\leq k}\mathcal{M}_{n_i}(\mathbb{C})\to\mathcal{M}_p(\mathbb{C})$ is a $*$-homomorphism iff there exist a $p\times p$ unitary $U$ and natural numbers $s_1,\ldots,s_k$ such that $\sum_{1\leq i\leq k}n_is_i=p$ and 
\[f(A_1,\dots,A_k)=U(A_1\otimes \Id_{s_1}\oplus \ldots \oplus A_k\otimes \Id_{s_k})U^*\]
(Here $\oplus$ denotes the block diagonal matrix.)
If moreover the $n_i$ are all non-zero, then the sequence $s_1\dots s_k$ is unique.
\end{proposition}

\begin{notation}
\begin{itemize}
\item Given a tuple of natural numbers $\widebar{n}:=(n_1,\ldots,n_k)$, let $\varphi_{\widebar{n}}:\bigoplus_i\mathcal{M}_{n_i}(\mathbb{C})\to \mathcal{M}_{\sum_i n_i}(\mathbb{C})$ be the canonical injection sending $(M_1,\ldots,M_k)$ to $M_1\oplus\ldots\oplus M_k)$.
\item Let $\Delta_{s,n}:\mathcal{M}_{n}(\CC)\to \mathcal{M}_{n}(\CC)^{\oplus s}$ be the duplication map given by the universal property of the product, which sends $A$ to $s$ copies of $A$. In particular $\Delta_{1,n}=\id_n$ and $\Delta_{0,n}=!_n$.
\item Let $\Ad_U:M\mapsto UMU^*$ be the adjoint operator.
\end{itemize}
For brevity and clarity of reading, we will often omit indices.
\end{notation}

Rephrasing the characterisation of Bratteli, we get:
\begin{corollary}
\label{cor:NF}
 $\Cstar\big([n_1,\ldots,n_k],[p]\big)$ consists exactly of the unital $*$-homomorphisms of the form $\Ad_U\circ \varphi_{\widebar{n}}\circ \bigoplus_i \Delta_{s_i,n_i}$ for some $p\times p$ unitary $U$ and $\sum_i s_in_i=p$.
\end{corollary}

\subsection{Completion of $\Unitary$}

The completion theorem \ref{thm:homo} is the main result of the section and rests upon a few lemmas:

\begin{lemma}
\label{lem:useful1}
$\Ad_U\circ\varphi\circ\Delta_{s,n}=\Ad_{U'}\circ\varphi\circ\Delta_{s,n}:\mathcal{M}_n\to\mathcal{M}_{ns}$ iff there exists an $s\times s$ unitary $V$ such that $U\circ (V\otimes \Id_n)=U'$. 
\end{lemma}

\begin{proof}
For the converse, notice that $\Delta_{s,n}=\Delta_{s,1}\otimes I_n$. Also note that $\varphi_{n,n}=\varphi_{1,1}\otimes Id_n$.
Then for all $U:ns\to ns,V:s\to s$ we have, using the fact that $1$ is the initial object:
\begin{align*}
&\Ad_U\circ \Ad_{V\otimes \Id_n}\circ\varphi\circ\Delta_{s,n}\\
&=\Ad_U\circ (\Ad_V \otimes \Id_n)\circ(\varphi\otimes \Id_n)\circ (\Delta_{s,1}\otimes \Id_n) \\
&= \Ad_U\circ \big((\Ad_V \circ \varphi \circ \Delta_{s,1})\otimes \Id_n\big) \\
&= \Ad_U \circ \big((\varphi \circ \Delta_{s,1})\otimes \Id_n\big) \\
&= \Ad_U \circ \varphi \circ \Delta_{s,n}
\end{align*}
We now prove the first part. Let $M:=\sum_{0\leq i<n}E_{j+im,j+im}$.
Then for all $k\notin\{j+im~|~i=0\dots n-1\}$ we have 
\[0=M_{k,k}=(UMU^*)_{k,k}=\sum_{0\leq i<n}a_{k,j+im}m_{j,j}\bar{a}_{k,j+im}\]
hence $\sum_{0\leq i<n}|a_{k,j+im}|^2=0$
so $\forall i\in\{0,\dots,n-1\}$, $a_{k,j+im}=0$.

For arbitrary $M$ and $(i,j)$ where $m_{i,j}$ is not necessarily $0$ we have:
\[m_{i,j}=\sum_{0\leq k<n}a_{i,i+km}m_{i,j}\bar{a}_{j,j+km}\] 
hence $\sum_{0\leq k<n}a_{i,i+km}\bar{a}_{j,j+km}=1$.

Using Cauchy-Schwarz inequality we get:
\begin{align*}
1&=\sum_{0\leq k<n}a_{i,i+km}\bar{a}_{j,j+km} \\
&=\langle(a_{i,i},\dots,a_{i,i+(n-1)m}),(a_{j,j},\dots,a_{j,j+(n-1)m})\rangle \\
&\leq \norm{(a_{i,i},\dots,a_{i,i+(n-1)m})}^2.\norm{(a_{j,j},\dots,a_{j,j+(n-1)m})}^2 \\
&= 1
\end{align*}
Again by Cauchy-Schwarz theorem, there exists $\lambda\in\mathbb{C}$ such that \[(a_{i,i},\dots,a_{i,i+(n-1)m})=\lambda (a_{j,j},\dots,a_{j,j+(n-1)m})\]
Replacing in the equality above leads to $1=\sum_{0\leq k<n}\bar{\lambda}|a_{i,i+km}|^2=\bar{\lambda}$ hence $\lambda=1$.
Consider $V=(a_{1+im,1+jm})_{0\leq i,j< n}$. It is a unitary matrix and it verifies $V\otimes I_m=U$.
\end{proof}

\begin{lemma}
\label{lem:useful2}
If $\Ad_U\circ\varphi_{\widebar{n}}\circ\bigoplus_i\Delta_{s_i,n_i}=\varphi_{\widebar{n}}\circ\bigoplus_i\Delta_{s_i,n_i}$, then $U=\bigoplus_i U_i$ for some $s_in_i\times s_in_i$ unitaries $U_i$.
\end{lemma}

\begin{proof}
It suffices to show that $U=U_1\oplus U_2$ with $U_1:s_1n_1\to s_1n_1$ and then conclude by induction on the size of $I$, the set of indices in the sum.

Let $M:=\sum_{s_1n_1<j\leq p} E_{j,j}$. 
Then, for $1\leq k\leq s_1n_1$, we obtain:
\[0=M_{k,k}=\sum_{s_1n_1<j\leq p} a_{k,j}M_{j,j}\bar{a_{k,j}}\]
Hence for all $1<k\leq s_1n_1,s_1n_1<j\leq p$ we have $a_{k,j}=0$.

As $U^*MU=M$, we get $UMU^*=M$ hence for all $1<k\leq s_1n_1,s_1n_1<j\leq p$, we have $a_{j,k}=0$. 
This indeed shows that $U=U_1\oplus U_2,$ as desired.
\end{proof}

\begin{theorem}
\label{thm:homo}
The functor $\Unitary\to(\Cstar)^{op}$ is a completion w.r.t. the embedding $
\RingoidTwo\otimes 1 + 0 ssss
\to
\RingoidTwo\otimes I \oplus N sscs$.
That is:
\[
\xymatrix{
\Unitary \ar[r]^{(\Embedding,\varphi)} \ar[dr]_{\forall (F,\psi)} 
&(\Cstar)^{op} \ar@{.>}[d]^{\exists!\widehat{F}} \\
& \forall\scatC 
}
\]
where $(F,\psi)\in\RingoidTwo\otimes \oone \oplus\ozero sscs$ and $\widehat{F}\in\RingoidTwo\otimes 1 + 0 ssss$.
\end{theorem}

\begin{proof}[Proof notes]
\underline{Uniqueness}: 
using corollary \ref{cor:NF} it's easier to show the categorical dual statement with in particular $\widehat{F}:\Cstar\to\scatC\opp$ and $\varphi,\psi$ being lax instead of colax, and conclude by a careful duality. 
$\widehat{F}$ has to strictly preserve $\otimes,\oplus$ and has to make the triangle commute so it must satisfy:
\begin{itemize}
\item $\widehat{F}([n_1,\ldots,n_k])=\bigoplus_i F(n_i)$
\item If $f=\Ad_{U}\circ\varphi\circ \bigoplus_i\Delta_i\in\Cstar$ then
$\widehat{F}(f)=F(U)\circ \psi\circ \bigoplus_i\Delta_i$
\end{itemize}
\underline{Existence}:
we define $\widehat{F}$ is the unique possible way and it is shown to be well defined by Lemmas \ref{lem:useful1} \& \ref{lem:useful2}. $\widehat{F}$ is proven to be a functor by a long but straightforward rewriting. It's easy to show that $\widehat{F}$ preserves $\oplus$ and $\otimes$ on morphisms.
\end{proof}

\section{Quantum channels}
\label{sec:cptp}

In this Section we prove our main theorem~(Thm.~\ref{thm:main}): von Neumann's model for full QM, $\CPTP$, is a canonical completion of the usual model of pure quantum theory with ancilla preparation, $\Isometry$. The main technical tool is Stinespring's theorem which characterises completely positive maps. 

\begin{notation}
	\begin{itemize}
		\item Given natural numbers $q\geq p$, let $R_{q,p}:\mathcal{M}_q(\mathbb{C})\to\mathcal{M}_p(\mathbb{C})$ denote the projection to the first $p$ rows and columns $A\mapsto A|_p$
		\item Given a category with an initial object $0$, we denote by $\subfac_A:0\to A$ the unique morphism to $A$
		\item Similarly, $!_A:A\to 1$ denotes the unique morphism from $A$ to the terminal object
	\end{itemize}
\end{notation}

\subsection{Stinespring's theorem}

In finite dimension and in the case of CPU maps the theorem can be stated as follows:

\begin{theorem}[Stinespring, e.g.~\cite{paulsen_2003}]
Let $p$ be a natural number. If $f:A\to \mathcal{M}_p(\mathbb{C})$ is a completely positive and unital map, then there is a natural number $q\geq p$ and a unital $*$-homomorphism $\pi$ making the following diagram commute:
\[
\xymatrix{
A \ar[dr]^{f} \ar@{.>}[d]_{\pi}
& 
\\
\mathcal{M}_q(\mathbb{C}) \ar[r]_{R_{q,p}} & \mathcal{M}_p(\mathbb{C})
}
\]
Moreover, $q$ can chosen to be the minimal such number: if $r \geq p$ and a $*$-homomorphism $h:A\to\mathcal{M}_r$ is such that $h(-)|_p=f(-)$ then $r\geq q$ and there is a unitary $U$ such that $\pi(-)=(Uh(-)U^*)|_q$. In a diagram: 
\[
\xymatrix{
& &
A \ar[dr]^{f} \ar[d]_{\pi} \ar@/_1pc/[dll]_{h}
& 
\\
\mathcal{M}_r(\mathbb{C}) \ar@/_2pc/[rrr]^{R_{r,p}} \ar@{.>}[r]_{\\Ad_U}
& \mathcal{M}_r(\mathbb{C}) \ar[r]_{R_{r,q}}
& \mathcal{M}_q(\mathbb{C}) \ar[r]_{R_{q,p}}
& \mathcal{M}_p(\mathbb{C})
}
\]
\end{theorem}

In other words, every CPU map can be written as a unital $*$-homomorphism followed by a restriction map, and possibly in a sort of minimal but non unique way. In categorical language this leads to the following:

\begin{corollary}
$\CPU\big([n_1,\ldots,n_k],[p]\big)$ are exactly morphisms of the form $R_{q,p}\circ \Ad_U\circ \varphi\circ \bigoplus_i \Delta_{s_i,n_i}$ for some $q\times q$ unitary $U$ and $\sum_i s_in_i=q\geq p$.
\end{corollary}

\begin{remark}
Recall that we have a functor $\Embedding:(\Isometry)^{op}\to\CPU$. We have $R_{r,q}=\Embedding(\id_q\oplus !_{r-q})$.
Because of this, $R_{r,q}$ is sometimes called purification (\cite{purification,ck-pictures}).
\end{remark}

\subsection{Completion of $\Isometry$}

We need a few key lemmas which are summed up in the proposition below, which gives essentially unique normal forms to the morphisms in $\CPTP$ as follows:

\begin{proposition}
\label{propn:useful}
$R_{m,p}\circ \Ad_U \circ \varphi \circ\bigoplus_{1\leq i \leq k}\Delta_{s_i,n_i}=R_{m,p}\circ \Ad_{U'} \circ \varphi \circ \bigoplus_{1\leq i \leq k}\Delta_{s_i,n_i}:\bigoplus_i\mathcal{M}_{n_i}\to\mathcal{M}_p$ iff there exist unitary matrices $P,Q_1,\dots, Q_k$ such that \[(\Id_{m-p}\oplus P)\circ U\circ \big((Q_1\otimes \Id_{n_1})\oplus\dots\oplus (Q_k\otimes \Id_{n_k})\big)=U'\]
\end{proposition}

\begin{proof}[Proof notes]
Suppose $f=\Ad_{V_1^*}\circ \pi_2=\Ad_{V_2^*}\circ \pi_2$ for unital $*$-homomorphisms $\pi_i$. There exists a minimal dilation $f=\Ad_{U^*}\circ\pi_1$. Using the corollary of the uniqueness in Stinespring's theorem, there exists two isometries $W_1,W_2$ such that $\Ad_{W_i}\circ \pi_1=\pi_2$ and $W_iU=V_i$. 

We now mimic the proof of Lemma \ref{lem:useful2} and we get that $W_i=\bigoplus_j W_{i,j}$. Indeed, one can write $\pi_2=\Ad_T\circ \varphi\circ\bigoplus_i\Delta_{s_i'}$ for some unitary $T$ and $\pi_1=\Ad_Q\circ \varphi\circ\bigoplus_i\Delta_{s_i}$ for some unitary $Q$. Without loss of generality we can take $\pi_2$ to be $\varphi\circ\bigoplus_i\Delta_{s_i'}$ and $\pi_1$ to be $\varphi\circ\bigoplus_i\Delta_{s_i}$, for instance by changing $W_i$ into $(T^*)^{-1}W_i(Q^*)^{-1}$. Let $W$ be either of $W_1$ or $W_2$.
Then, as in the proof of Lemma \ref{lem:useful2}, we get that 
\[m_{i,j}=\sum_{1\leq p\leq n}\sum_{1\leq k\leq n} w_{i,k}m_{k,p}\bar{w}_{j,p}\]
Taking $M$ to be ones on the diagonal after the first block (of size $s_1n_1$) and $0$ elsewhere gives that for all $1\leq i\leq s_1n_1$, $\sum_{n_1s_1< p\leq n}w_{i,p}\bar{w}_{i,p}=0$ hence $w_{i,p}=0$ for all $1\leq i\leq s_1n_1$, and for every $n_1s_1< p\leq n$. Doing this for the other blocks as well shows that $W$ is of the form $W=\bigoplus_i W_i$ as desired.

Now precomposing by the injection $\mathcal{M}_{n_j}\to \bigoplus_i\mathcal{M}_{n_i}$ and using the known result for quantum channels (see for instance \cite{wolf}) and the characterisation of isometries (Proposition \ref{propn:caracIsometry}), 
$W_{i,j}=(P_{i,j}\otimes Id)\circ (Id\oplus \subfac)$. 

Putting everything together, there is a permutation $\widetilde{\gamma}$ independant of $i$ such that $W_i= (P_{i,1}\otimes Id \oplus \dots \oplus P_{i,k}\otimes Id)\circ \widetilde{\gamma} \circ (Id \oplus \subfac)$. Hence 
\begin{align*}
&(P_{1,1}P_{2,1}^{-1}\otimes Id \oplus\dots\oplus P_{1,k}P_{2,k}^{-1}\otimes Id)\circ V_2 \\
&=(P_{1,1}P_{2,1}^{-1}\otimes Id\oplus \dots\oplus P_{1,k}P_{2,k}^{-1}\otimes Id)\circ W_2U \\
&=(P_{1,1}P_{2,1}^{-1}\otimes Id \oplus\dots\oplus P_{1,k}P_{2,k}^{-1}\otimes Id)\\
&\qquad\circ (P_{2,1}\otimes Id \oplus\dots\oplus P_{2,k}\otimes Id)\circ \widetilde{\gamma} \circ (Id \oplus \subfac) \circ U \\
&= (P_{1,1}\otimes Id \oplus\dots\oplus P_{1,k}\otimes Id)\circ \widetilde{\gamma} \circ (Id \oplus \subfac) \circ U \\
&= W_1 \circ U = V_1 
\end{align*}
and we conclude by Proposition \ref{propn:caracIsometry}.
\end{proof}

\begin{theorem}
\label{thm:main}
The functor $\Isometry\to\CPTP$ is a completion w.r.t. the embedding
$
\RingoidTwo\otimes 1 + 0 ssss
\to
\RingoidTwo\otimes \oone \oplus 0 sscs$.
That is:
\[\xymatrix{
\Isometry \ar[r]^(0.55){(\Embedding,\varphi)} \ar[dr]_{\forall (F,\psi)}
& \CPTP \ar@{.>}[d]^{\exists!\widehat{F}} \\
& \forall\scatC
}
\]
where $(F,\psi)\in\RingoidTwo\otimes 1 \oplus\ozero sscs$ and $\widehat{F}\in\RingoidTwo\otimes 1 + 0 ssss$.
\end{theorem}

\begin{proof}[Proof notes]
It's not hard to show uniqueness using Stinespring's theorem and the fact that $\widehat{F}$ has to be a functor and has to preserve $\oplus,\otimes$.\\
Existence: there is a unique way to define $\widehat{F}$ on objects. $\widehat{F}$ is defined on morphisms essentially by using Proposition \ref{propn:useful}. Note that it's important that we don't define $\widehat{F}$ on minimal dilations only as the composition of two minimal dilations does not trivially reduce to a minimal dilation. Again using Proposition \ref{propn:useful}, one can show that $\widehat{F}$ preserves composition. It is then easy to show that $\widehat{F}$ preserves $\oplus$ and $\otimes$.
\end{proof}

\begin{remark}
We have the following picture:
\[\xymatrix{
(\Unitary)^{op} \ar[r] \ar[d] & \Cstar \ar@{.>}[d] \\
(\Isometry)^{op} \ar[r] & \CPU
}
\]
Both horizontal arrows express the same completion, and the left vertical arrow is a different completion. The dotted arrow is then given by the universal property of $\Cstar$ as a completion of $\Unitary$ (Thm.~\ref{thm:homo}). Using Theorem \ref{thm:unitary}, this sheds a new light on Stinespring's theorem which 
can now be understood as the lifting to the operator-algebra level of the completion of unitaries into isometries~(\S\ref{subsec:C}). 
Intuitively, what makes the dotted functor non-full is the image by the bottom-horizontal completion of what is added by the left vertical one, and Stinespring's theorem makes this precise.
\end{remark}

\begin{remark}
It is perhaps perplexing that ever since Notation~\ref{ringoid_not}
we have considered bipermutative functors that are colax with respect to $\oplus$ but strict for $N$. 
However, for any $\oplus$-colax bipermutative functor, 
when $N$ is initial and $F(I)$ is terminal, we have that $F(N)$ is necessarily also initial,
because for any object $A$ there is a morphism
$F(N) \xrightarrow \psi N \xrightarrow{!} A$.
(To show that this morphism is unique we must use the terminality of $F(I)$).
So in this situation, colax for $N$ implies $F(N)\cong N$. \end{remark}

\section{Completion as topologically enriched categories}
\label{sec:enriched}

We now extend our algebraic framework and consider the context of topologically-enriched category theory. The main goal is to show analogues of Theorems \ref{thm:homo} \& \ref{thm:main} in the topologically-enriched setting. This means that the norm topology on unital $*$-homomorphisms is the canonical one induced by the norm topology on unitaries and that the norm topology on CPTP maps is the canonical one induced by the norm topology on isometries.

The basic theory of enriched category theory is for instance covered in \cite{kelly1982basic}.

\subsection{Topologically enriched bipermutative categories}


\begin{definition}
	Given a symmetric monoidal category $\catV$, a $\catV$-category $\scatC$ is given by the following:
	\begin{itemize}
		\item a set of objects $Obj(\scatC)$
		\item for each pair of objects $(a,b)$ of $\scatC$ an object $\scatC(a,b)\in Obj(\catV)$
		\item for each triple $(a,b,c)$ of objects of $\scatC$ a morphism $\circ_{a,b,c}:\scatC(b,c)\otimes \scatC(a,b)\to \scatC(a,c)$ in $\catV$ -- called the composition morphism
		\item for each object $a\in Obj(\scatC)$ a morphism $j_a:I\to \scatC(a,a)$ in $\catV$ -- called the identity morphism
such that composition is unital and associative.
	\end{itemize}
\end{definition}

\begin{example}
Every (locally small) category is $\Set$-enriched, where $\Set$ is the category of sets and functions, seen as a Cartesian monoidal category.
\end{example}

Given a monoidal category $\catV$, a $\catV$-bipermutative category is a category $\scatC$ enriched over $\mathcal{V}$ with $\mathcal{V}$-enriched bifunctors $\oplus,\otimes:\scatC\times\scatC\to\scatC$ and $\catV$-enriched natural isomorphisms $\gamma,\gamma',\delta^\sharp$ such that the coherence conditions of bipermutative categories are satisfied. There is a category of $\catV$-bipermutative categories and $\catV$-bipermutative functors between them.

Here, we consider $\catV$ to be the Cartesian monoidal category $(\Topo,\times)$ of topological spaces and continuous maps equipped with the Cartesian product as monoidal tensor product. 
We note that early applications of bipermutative categories~\cite{may1977ring} also make use of topological categories and topological enrichment, although those applications in algebraic topology are different from ours. 

 If we equip $\CC^n$ with the Euclidean norm $\Vert(x_1,\ldots,x_n)\Vert:=\sum_i|x_i|^2$ then we can equip matrix spaces $\mathcal{M}_{n}(\CC)$ with the spectral norm $\norm{-}_2$ which is defined as $\Vert M\Vert_2:=sup_{\Vert u\Vert=1}\Vert Mu\Vert$. 

Every norm $\norm{-}$ on a space $M$ induces a topology on $M$ where opens are given as follows. For every $x\in M$ and every non negative real $r$ the set $\{y\in M~|~\norm{y-x}<r\}$ is open. Then an arbitrary open is given by arbitrary unions and finite intersections from such sets. Hence we can see the homsets in $\Unitary$ and in $\Isometry$ as topological spaces whose topology is induced by the spectral norm.

Morphisms in $\CPTP$, $\CPU$, $\Cstar$ can be equipped with the operator norm $\norm{-}_{op}$ defined as $\norm{f}_{op}:= sup_{\norm{a}_2=1}\norm{f(a)}_2$. The homsets in these categories can thus also be seen as topological spaces whose topology is induced by the operator norm. The topology gives a lot of information on the space. For instance $\CPTP\big([1],[1,1]\big)$ is homeomorphic to the unit interval $[0,1]$ and is to been understood as a probability. $\CPTP\big([1],[2]\big)$ can be understood as the state space for qubits, the Bloch sphere, and is homeomorphic to the 3-ball. On the other hand, $\Cstar\opp\big([2],[1,1]\big)$ is homeomorphic to the 2-sphere plus two points: computationally, given a qubit, we must either measure it on some axis of the Bloch 2-sphere, or discard it and return a classical bit ($0$ or $1$).
\hide{More on quantum topology can be found in \cite{kauffman-baadhio}.}


\begin{lemma}
\label{lem:enriched}
With the topology given above, $\Unitary$, $\Isometry$, $\CPTP$, $\CPU$, $\Cstar$ are $\Topo$-enriched bipermutative categories. That is for all objects $A,B,C$ the composition map $\circ_{A,B,C}:\scatC(B,C)\otimes \scatC(A,B)\to \scatC(A,C)$, the first tensor product $\big(A\otimes(-)\big)_{B,C}:\scatC(B,C)\to \scatC(A\otimes B,A\otimes C)$, and the second tensor product $\big(A\oplus(-)\big)_{B,C}:\scatC(B,C)\to \scatC(A\oplus B,A\oplus C)$ are continuous maps.
\end{lemma}

\begin{notation}
\label{not:enriched}
Similarly to Notation \ref{ringoid_not} we denote by 
\begin{itemize}
	\item $\TopRingoidTwo\otimes I \oplus N sscs$ the category of $\Topo$-bipermutative categories and $\oplus$-colax $\Topo$-bipermutative functors
	\item $\TopRingoidTwo\otimes I \oplus 0 sscs$ for the category of $\Topo$-bipermutative categories for which $N$ is an initial object and $\oplus$-colax $\Topo$-bipermutative functors
	\item $\TopRingoidTwo\otimes 1 +0ssss$ for the category of $\Topo$-bipermutative categories for which $\oplus$ is a coproduct and $I$ is terminal, and strict $\Topo$-bipermutative morphisms between them
	\item $\Topo$-$\scatC$ any of the categories $\scatC$ from Lemma \ref{lem:enriched} seen as $\Topo$-enriched
\end{itemize}
where all the bipermutative functors actually strictly preserve the unit $N$ of $\oplus$.
\end{notation}

\subsection{Topologically enriched completion for $\Cstar$}

The main lemma we use is proved in \cite[Lemma 3.7]{amini2014category}, 
which requires an important lemma from Glimm \cite{glimm1960certain}:

\begin{lemma}[\cite{amini2014category}]
Let $\phi,\psi:A\to B$ be unital $*$-homomorphisms between finite dimensional C*-algebras such that $\norm{\phi-\psi}_{op}<1$. Then there is a unitary $U$ in $B$ such that $\phi=\Ad_U\circ \psi$.
\end{lemma}

\begin{definition}
Given $n,\widebar{m}:=(m_1,\ldots,m_k)$, a \emph{$(n,\widebar{m})$-Bratteli tuple $(s_1,\ldots,s_k)$} is a tuple of natural numbers such that $\sum_i s_im_i=n$.
The set of $(n,\widebar{m})$-Bratteli tuples is denoted by $B_n^{\widebar{m}}$.
Given $(s_1,\ldots,s_k)\in B_n^{\widebar{m}}$ we define a group
\[G_{\widebar{s},\widebar{m},n}:=\{u_1\otimes \Id_{s_1}\oplus\ldots u_k\otimes \Id_{s_k}: u_i\in\mathbb{U}(m_i)\}\subset\UU(n)\]
Given a lax bipermutative functor $(F,\psi):\Unitary\to\scatC$  and $\widebar{s}:=(s_1,\ldots,s_k)\in B_n^{\widebar{m}}$, there is a canonical map that sends $F(U)$ to $F(U)\circ\psi\circ\bigoplus_i\Delta_{s_i,m_i}$ given by precomposition, which we denote by $can_{\widebar{s},\widebar{m},n}:[Fn,Fn]\to [\bigoplus_i Fm_i,Fn]$. 
\end{definition}

$G_{\widebar{s},\widebar{m},n}$ is a subgroup of $\UU(n)$ and acts on it by right multiplication and we can thus consider the quotient $\UU(n)/G_{\widebar{s},\widebar{m},n}$. See for instance \cite{dummit2004abstract} for more on quotients by an action of a group.
One can show that $can_{\widebar{s},\widebar{m},n}\circ F_{n,n}:\Unitary(n,n)\to \scatC\big(\bigoplus_i Fn_i,Fn\big)$ respects the quotient by $G_{\widebar{s},\widebar{m},n}$.

The following lemma is the key point to show that the lifted bipermutative morphism -- written $\widehat{F}$ in Sections \ref{sec:homo} \& \ref{sec:cptp} -- is continuous whenever the colax one $(F,\psi)$ is:

\begin{lemma}
\label{lem:canonical}
The following spaces are homeomorphic:
\[\Cstar\big([m_1,\ldots,m_k],[n]\big)\cong \coprod_{\widebar{s}\in B_n^{\widebar{m}}}\mathbb{U}(n)/G_{\widebar{s},\widebar{m},n}\]
\end{lemma}

The lemma implies that $\Cstar\big([m_1,\ldots,m_k],[n]\big)$ has a connected component for each Bratteli tuple in $B_n^{\widebar{m}}$. Each of these connected components is a smooth manifold -- a space locally homeomorphic to an Euclidean space -- but they usually have different dimensions. For instance we recover that $\Cstar([1,1],[2])$ has 3 connected components, two of which are points so 0-dimensional and the other one is the 2-sphere.

With those ingredients at hand we show an analogue of Theorem \ref{thm:homo} in the $\Topo$-enriched setting. Informally, this says that topology on unital $*$-homomorphisms is the canonical one induced by the one on unitaries.

\begin{theorem}
\label{thm:homoEnriched}
The functor $\Topo$-$\Unitary\to(\Topo$-$\Cstar)^{op}$ is a completion w.r.t. the embedding 
$\TopRingoidTwo\otimes 1 + 0 ssss
\to
\TopRingoidTwo\otimes I \oplus N sscs$. 
That is:
\[
\xymatrix{
\Topo\text{-}\Unitary \ar[r]^{(\Embedding,\varphi)} \ar[dr]_{\forall (F,\psi)} 
&(\Topo\text{-}\Cstar)^{op} \ar@{.>}[d]^{\exists!\widehat{F}} \\
& \forall\scatC 
}
\]
where $(F,\psi)\in\TopRingoidTwo\otimes I \oplus N sscs$ and $\widehat{F}\in\TopRingoidTwo\otimes 1 + 0 ssss$. See Notation \ref{not:enriched}.
\end{theorem}

Theorems \ref{thm:homo} \& \ref{thm:homoEnriched} then combined say that for every function $F_{n,n}:\Unitary(n,n)\to [Fn,Fn]$ there is a unique lifted function $\widehat{F}_{n,n}:\Cstar^{op}([n],[n])\to [Fn,Fn]$, which is moreover continuous whenever $F_{n,n}$ is.

\begin{proof}[Proof notes]
From Theorem \ref{thm:homo} uniqueness already holds for the $\Set$ version. 
The existence of a $\Set$ function is again ensured by Theorem \ref{thm:homo}. It remains to show continuity. $\widehat{F}_{n,n}$ is shown to be continuous by Lemma \ref{lem:canonical} and the universal property of the quotient topology. Then, all morphisms $\widehat{F}_{A,B}$ are constructed from these, the canonical maps and the universal property of the coproduct. The resulting morphisms are all continuous by construction.
\end{proof}

\subsection{Topologically enriched completion for $\CPTP$}

Finally, we prove that the norm topology on CPTP maps is the canonical one induced by the norm topology on isometries. In particular, starting from isometries rather than unitaries forces a lot of differences to the topology: as opposed to the case of $\Cstar$, the homsets now only have one connected component. 

There is no maximal Stinespring dilation for a CPU map, but it is always possible to chose a dilation space which is common to all CPU maps $\bigoplus_i\mathcal{M}_{m_i}(\CC)\to\mathcal{M}_n(\CC)$:

\begin{lemma}[e.g. \cite{paulsen_2003}, p.50]
Any morphism $f\in[\bigoplus_i\mathcal{M}_{m_i}(\CC),\mathcal{M}_n(\CC)]$ admits a Stinespring dilation in $\mathcal{M}_D(\CC)$,  where $D:=\sum_i m^2_in$.
\end{lemma}

This implies that as opposed to the case of unitaries, we can embed all the information needed for CPU maps $\bigoplus_i\mathcal{M}_{m_i}(\CC)\to\mathcal{M}_n(\CC)$ in a single isometry space. This leads us to:

\begin{theorem}
\label{thm:cptpEnriched} 
The functor $\Topo$-$\Isometry\to\Topo$-$\CPTP$ is a completion w.r.t. the embedding 
$\TopRingoidTwo\otimes 1 + 0 ssss
\to
\TopRingoidTwo\otimes \oone \oplus 0 sscs$.
That is:
\[\xymatrix{
\Topo\text{-}\Isometry \ar[r]^(0.55){(\Embedding,\varphi)} \ar[dr]_{\forall (F,\psi)}
& \Topo\text{-}\CPTP \ar@{.>}[d]^{\exists!\widehat{F}} \\
& \forall\scatC
}
\]
where $(F,\psi)\in\TopRingoidTwo\otimes \oone \oplus 0 sscs$ and $\widehat{F}\in\TopRingoidTwo\otimes 1 + 0 ssss$. See Notation \ref{not:enriched}.
\end{theorem}

\begin{proof}[Proof notes]
We prove the dual result. We proceed similarly to the proof of Theorem \ref{thm:homoEnriched}. The base case is obtained by the universal property of the quotient of the topological space of dilations $\Isometry(n,D)\to \CPU\big([m_1,\ldots,m_k],[n]\big)$ where $D=\sum_i m^2_in$. 
\end{proof}

Isometries can be shown to be a connected smooth manifold, hence as a quotient of such a space, $\CPTP$ homsets are connected. Operationally, the isometry $\subfac:0\to1$ induces a map $\Embedding(\id_1\oplus\subfac):\CC\to\mathcal{M}_2(\CC)$ which is to be understood as a bridge between the classical and quantum world as it turns a bit 0 into a qubit in state $\ket{0}$. $\Cstar\opp\big([2],[1,1]\big)$ for instance has three connected components: two of them are points to be understood as bits and then the 2-sphere of pure qubits, so the bits are not connected to specific points on the sphere. 

\hide{\mh{discussion somewhere on the fact that adding discarding when there a bridge between the classical and the quantum world somehow forces to keep one connected component, which was not true in $\Cstar$, maybe a comment somewhere on the fact that isometries are connected (as path connected) just like unitaries, which implies CPTP maps are path connected as it's a quotient space.}
}

\section{Context}
\label{sec:context}
We conclude this paper by mentioning connections with related programmes. 

\subsection*{Brief comparison with the CPM construction}

Categorical quantum mechanics (CQM) is a successful abstraction of the compact closed category of natural numbers and \emph{all} complex matrices by using `dagger' structures (\cite{cqm},\cite{ck-pictures}). Since this category has biproducts and is compact closed, it is a rig category (e.g.~\cite{cqm}, Prop.~5.3, although rig categories and bipermutative categories have not been considered in CQM in full generality). 
Finite dimensional C*-algebras are described abstractly as Frobenius structures, through the CPM construction~(\cite{coecke-cp},\cite{selinger2007dagger}), and variants thereof.
This has been described by a universal property~\cite{ch-ax-cp}, and to that extent is related to our characterisation, although it is unclear how that universal property is related to ours. In particular,  coproducts and additive monoidal structure ($\oplus$) typically do not play an explicit role in~\cite{ch-ax-cp}, in contrast to our work, where they are central to our understanding of quantum control and classical data. 

A broader difference between our work and the CQM approach is that the CQM-style work typically begins with \emph{all} linear maps and CPM provides \emph{all} completely positive maps, not only isometries and trace-preserving maps. Within the category of all completely positive maps, one can cut down to the trace-preserving ones by imposing causality conditions, formulated using the dagger in an elegant categorical way. An advantage of this approach is that the non-trace-preserving maps expose further categorical structure, e.g. for instance one can consider free biproduct completions (\cite{selinger2007dagger}, \S5) by taking advantage of the additional additive structure of the hom-sets. 
\emph{However,} although they are useful for calculations, the non-trace-preserving completely positive maps don't have a real physical meaning. Our philosophy is to not consider them at all when using universal properties to derive theories of physics. In this way our approach is different in spirit from CQM using the CPM construction.


\subsection*{Finite / infinite dimensions}
Our work here is focused on finite dimensional systems, as is common in quantum computation and information theory. More broadly in quantum physics, infinite dimensional systems are often considered. The categorical axiomatics of these systems is in its infancy, but some recent steps have been made (e.g.~\cite{ah-hstar,dagger-mix,rennela2017infinite,ww-paschke}).

\subsection*{Topology and metrics}
We motivated our work on topological enrichment (\S\ref{sec:enriched}) in terms of other theoretical work on quantum topology (\S\ref{sec:topology}). 
There are many more practical applications of quantum information that use notions of convergence, but there they use metrics rather than topology. 
There can be many metrics that generate the same topology, and indeed there are many metrics at play in 
quantum information (see \cite{Nielsen:2011:QCQ:1972505}, \S9, for an overview). 
It is possible that our Theorem~\ref{thm:cptpEnriched}, giving a universal property for the topological enrichment, could be refined to give a universal property to some metric on CPTP maps. 
This is a subtle point, as is elaborated in work (\cite{cts-stinespring},\cite{kretschmann2008information}) on the metrics involved in a continuous Stinespring's theorem. (These papers served as an inspiration for our topological result.) 

We note that, away from the quantum area, universal properties have recently been used to characterize metrics (e.g.~\cite{mardare2016quantitative,mardare2018axiomatizability}), and this may serve as an inspiration for future work.

\hide{

There is a now long tradition of using category theory as a tool for investigating foundations of quantum theories \cite{ch-ax-cp,cpstar,ch-purity}, using presheaf categories as in \cite{rennela2017infinite, rennela2015complete, staton2018effect, malherbe2013presheaf} and for the study of non-locality and contextuality \cite{abramsky2011sheaf}. 
Our work here is more focused around Stinespring theorem in finite dimension and is in a way closer to \cite{ww-paschke} in infinite dimension where Stinespring theorem is found to satisfy a universal property. 

On the other hand, our work on metric spaces provides a justification for the metric on super-operators and has to do with the theoretical investigation on the approximation of unitary gates and fault tolerence, in the vein of diagramatic calculus such as ZX \cite{backens2014zx, jeandel2018complete, ng2017universal}.}

\subsection*{Quantum programming languages}
The language of bipermutative categories can be thought of as a prototypical calculus for programming
circuits that allows control. This has been proposed earlier as a language for quantum circuits~\cite{green-altenkirch}. Selinger's QPL~\cite{selinger2004towards} and Adam's QPEL~\cite{adams-qpel} involve the related notion of distributive monoidal categories, which also play a role in Tull's categorical analysis \cite{tull-otp} of operational probabilistic theories through monoidal effectuses (\cite{cjww-effectus}, \S10). All this work has been a source of inspiration in our development.

In some ways a universal property is an equational characterization of a structure, and there have been other equational characterizations of quantum programming, including the measurement calculus~\cite{measurement-calculus}, the ZX calculus~(\cite{backens2014zx},\cite{jeandel2018complete},\cite{ng2017universal}), and the work by the second author in~\cite{staton2015algebraic}. These are more syntactic than the present work, which can be an advantage, but an advantage of the present work is that it is arguably more canonical through its categorical nature. 
We highlight in particular the recent extention of ZX with hiding~\cite{carette2019completeness}, 
which indeed makes an explicit connection to the universal properties considered in~\cite{huotuniversal}.

\hide{
\subsection*{Understanding probabilistic theories}
There has been a lot of recent work on understanding probabilities in and with algebraic theories \cite{mardare2016quantitative, mardare2018axiomatizability, staton2018beta}. In particular, relating probabilities from quantum and usual probability theory has attracted attention \cite{jacobs2016expectation,furber2013kleisli} as well as some reconstruction of quantum theory in generalized probabilistic theories \cite{purification,chiribella2011informational}. Our purely categorical account will hopefully lead to some insight in these directions.
}

\subsection*{Overall summary}

We have proposed the categorical framework of bipermutative categories in Section \textsection\ref{sec:prel} to study the connection between pure and mixed states. We proved in Section \textsection\ref{sec:homo} that the category of unital $*$-homomorphisms is a canonical completion of the category of unitaries and we used this result to prove in Section \textsection\ref{sec:cptp} that the category of completely positive trace-preserving maps is a canonical completion of the category of isometries.
We then went beyond the discrete framework in Section \textsection\ref{sec:enriched} and considered topologically enriched categories where we proved that the norm topology of CPTP maps is the canonical topology induced by the norm topology on isometries.


\section*{Acknowledgment}
We would like to thank the Quantum and PL groups from Oxford and the Quantum group from Nijmegen for useful discussions. We are also grateful to the authors of the Tikz and Qasm2circ packages and to the anonymous reviewers for their comments which improved this paper. We are grateful for all the feedback at QPL about our work.

\hide{We published a preliminary result in this direction in the proceedings of QPL 2018~\cite{huotuniversal}. That paper is not about bipermutative categories but merely symmetric monoidal categories with a terminal unit. We are grateful for all the feedback at QPL.} 

Research supported by a Royal Society Fellowship and Enhancement Award, and ESPRC Grant EP/N007387/1. 

\newpage


\bibliographystyle{IEEEtranS} 
\bibliography{licsbib.bib}
\end{document}